\documentclass[useAMS,usenatbib,usegraphicx]{mn2e}

\usepackage[latin1]{inputenc}
\usepackage{amsmath,amsfonts,amssymb}
\usepackage{color}
\usepackage{times}

\newcommand{\Msun}{\ensuremath{\textrm{M}_{\odot}}}

\newcommand{\kms}{km\hspace{0.25em}s$^{-1}$}

\newcommand{\OI}{\mbox{O\hspace{0.25em}{\sc i}}}
\newcommand{\OII}{\mbox{O\hspace{0.25em}{\sc ii}}}

\newcommand{\NaI}{\mbox{Na\hspace{0.25em}{\sc i}}}

\newcommand{\MgI}{\mbox{Mg\hspace{0.25em}{\sc i}}}

\newcommand{\SI}{\mbox{S\hspace{0.25em}{\sc i}}}

\newcommand{\SiI}{\mbox{Si\hspace{0.25em}{\sc i}}}

\newcommand{\CaII}{\mbox{Ca\hspace{0.25em}{\sc ii}}}

\newcommand{\FeII}{\mbox{Fe\hspace{0.25em}{\sc ii}}}
\newcommand{\FeIII}{\mbox{Fe\hspace{0.25em}{\sc iii}}}

\newcommand{\Cofs}{$^{56}$Co}
\newcommand{\Nifs}{$^{56}$Ni}

\newcommand{\Mej}{M$_{\textrm{ej}}$}

\newcommand{\KE}{E$_{\rm K}$}

\newcommand{\KEPLER}{\ensuremath{\mathrm{\texttt{KEPLER}}}}

\newcommand{\aap}{A\&A}

\newcommand{\apj}{ApJ}

\newcommand{\apjl}{ApJ}

\newcommand{\mnras}{MNRAS}

\newcommand{\nat}{Nature}

\newcommand{\prl}{Phys. Rev. Lett.}

\newcommand{\sci}{Sci}

\newcommand{\araa}{ARA\&A}

\newcommand{\eg}{e.g.,\ }
\newcommand{\ie}{i.e.,\ }

\def\gsim{\mathrel{\rlap{\lower 4pt \hbox{\hskip 1pt $\sim$}}\raise 1pt \hbox {$>$}}}
\def\lsim{\mathrel{\rlap{\lower 4pt \hbox{\hskip 1pt $\sim$}}\raise 1pt \hbox {$<$}}}
\def\gtaprx {\lower .1ex\hbox{\rlap{\raise .6ex\hbox{\hskip .3ex
	{\ifmmode{\scriptscriptstyle >}\else
		{$\scriptscriptstyle >$}\fi}}}
	\kern -.4ex{\ifmmode{\scriptscriptstyle \sim}\else
		{$\scriptscriptstyle\sim$}\fi}}}
\def\ltaprx {\lower .1ex\hbox{\rlap{\raise .6ex\hbox{\hskip .3ex
	{\ifmmode{\scriptscriptstyle <}\else
		{$\scriptscriptstyle <$}\fi}}}
	\kern -.4ex{\ifmmode{\scriptscriptstyle \sim}\else
		{$\scriptscriptstyle\sim$}\fi}}}

\definecolor{myorange}{rgb}{0.9,0.6,0.0}

\definecolor{myredder}{rgb}{1.0,0.0,0.0}

\definecolor{mygreen}{rgb}{0.0,0.7,0.0}

\definecolor{myblue}{rgb}{0.0,0.0,1.0}

\definecolor{myyellow}{rgb}{0.92,0.89,0.00}

\begin{document}

\title[The nature of PISN candidates]
  {The nature of PISN candidates: clues from nebular spectra}
\author[P. A.~Mazzali et al.]{P. A. Mazzali$^{1,2,3}$
\thanks{E-mail: P.Mazzali@ljmu.ac.uk}, 
T. J. Moriya$^3$, M. Tanaka$^{3,4}$, and S. E. Woosley$^5$\\
\\
  $^1$Astrophysics Research Institute, Liverpool John Moores University, IC2, 
  Liverpool Science Park, 146 Brownlow Hill, Liverpool L3 5RF, UK\\
  $^2$Max-Planck-Institut f\"ur Astrophysik, Karl-Schwarzschild Str. 1, 
  D-85748 Garching, Germany\\
  $^3$Division of Theoretical Astronomy, National Astronomical Observatory of 
  Japan, National Institutes of Natural Sciences, \\
2-21-1 Osawa, Mitaka, Tokyo 181-8588, Japan \\
  $^4$Astronomical Institute, Tohoku University, 6-3 Aramaki Aza-Aoba, Aoba-ku, 
  Sendai 980-8578, Japan \\
  $^5$Department of Astronomy and Astrophysics, University of California, 
  Santa Cruz, CA 95064, USA
}

\date{Accepted ... Received ...; in original form ...}
\pubyear{2019}
\volume{}
\pagerange{}

\maketitle

\begin{abstract}  
A group of super-luminous supernovae (SL-SNe) characterised by broad light
curves have been suggested to be Pair Instability SNe (PISNe). Nebular spectra
computed using PISN models have failed to reproduce the broad emission lines
observed in these SNe, casting doubts on their true nature. Here, models of both
PISNe and the explosion following the collapse of the core of a very massive
star ($100$\,\Msun) are used to compute nebular spectra, which are compared to
the spectrum of the prototypical PISN candidate, SN\,2007bi. PISN models are
confirmed to produce synthetic spectra showing narrow emission lines, resulting
from the confinement of \Nifs\ to the lowest velocities ($\lsim 2000$\,\kms) and
in clear disagreement with the spectrum of SN\,2007bi. Spectra more closely
resembling SN\,2007bi are obtained if the PISN models are fully mixed in
abundance.  Massive core-collapse models produce enough \Nifs\ to power the
light curve of PISN candidates, but their spectra are also not adequate. The
nebular spectrum of SN\,2007bi can be successfully reproduced if the inner
region is artificially filled with oxygen-rich, low-velocity ejecta. This most
likely requires a grossly aspherical explosion. A major difference between PISN
and massive collapse models is that the former emit much more strongly in the
NIR. It is concluded that: a) current PISN candidates, in particular SN\,2007bi,
are more likely the result of the collapse and explosion of massive stars below
the PI limit; b) significant asymmetry is required to reproduce the late-time
spectrum of SN\,2007bi.  
\end{abstract}


\section{Introduction}
\label{sec:introduction}

The class of SLSNe has been defined to include SN events that reach $M<-21$\,mag
\citep{galyam2012}. This paper divided SLSNe into three spectro-photometric
sub-classes. 

SLSNe-II show strong H lines in emission, and are likely to be the result of
the interaction between energetic SN ejecta and a massive, extended, H-rich
circum-stellar medium (CSM), just like SNe\,IIn, which are only different by
reaching a lower luminosity \citep[for a primer on SN classification,
see][]{filipp97}. 

Other SLSNe have Type\,Ib/c spectra, and were divided by \citet{galyam2012} into
two groups. One, by far the dominant one by number, called SLSN-I, includes SNe
that show absorption lines of heavy and intermediate-mass (IME) elements,
similar to SNe\,Ic \citep{pasto10}. Recently, a more accurate determination of
the luminosity above which SNe start to show SLSN characteristics has been
determined as $-19.8$\,mags \citep{decia18,quimby18}. Their evolution follows a
common pattern \citep{nicholl13}, and the most unusual feature is the presence
of a set of strong \OII\ lines in absorption at blue wavelengths before peak.
These lines arise from very highly excited levels, which require a
non-equilibrium population such as could be achieved if non-thermal
ionization/excitation processes are active \citep{mazzali16}. Such non-thermal
processes may be due to deposition of radioactive decay products as in SNe\,Ib
\citep{lucy91}, but the higher energetic requirements for \OII\ and the
unrealistically large \Nifs\ mass required to fit the light curves of some
SLSNe-I suggest that X-rays from the shock induced in the ejecta by the
interaction with a magnetar-driven wind may be responsible for the
overexcitation of \OII\ \citep{mazzali16}. Magnetar wind impact models are
commonly used to reproduce SLSN-I light curves as an alternative to radioactive
driving: interaction with a magnetar-driven wind could explain the high
luminosity of SLSNe-I, some of which seem to decline rather sharply some time
after peak, making it difficult to explain the high luminosity as due entirely
to \Nifs. Models of SLSN-I light curves based on the magnetar impact model have
typically assumed a very small mass of \Nifs\
\citep{woosley2010,kasenbildsten2010}. \citet{mazzali16} found that SLSNe-I may
have a large range of ejected masses, but are typically characterised by a value
of the ratio of explosion kinetic energy (\KE) and ejected mass (\Mej) close to
\KE/\Mej$= 1$ [$10^{51}$\,erg/\Msun]. They also claimed that He may be present
is some SLSNe-I ejecta. This would be supported also by the detection of H
emission lines due to CSM impact in at least three cases \citep{yan2017}.
Interestingly, a luminous SN, which shares some spectroscopic properties with
SLSNe-I although it did not reach $M<-21$\,mag and it had higher $E/M$, was
found in association with one of the few known cases of Ultra-Long Gamma-ray
Bursts (GRB), and the only case for which spectroscopic observation of a SN bump
were possible \citep{levan2014,greiner15}.

This leaves us with the smallest group of SLSNe, which \citet{galyam2012}
defined as SLSN-R, for radioactive.  This group includes only a small number of
SNe, notably SN\,2007bi, SN\,1999as, the first SLSN observed
\citep{knop1999iauc}, PTF12dam  \citep{nicholl13}, SN\,2015bn
\citep{jerk16,nicholl15}, and PS1-14bj \citep{lunnan2016}. While also showing
SN\,Ic spectra, these SNe decline slowly in luminosity, following the \Cofs\
decline rate, and are not known to show \OII\ lines early on (but early
observations are scarce and the rise time is highly uncertain). Their high
luminosity implies a large \Nifs\ mass (typically a few \Msun), and their
extended light curves lead to estimates of the ejecta mass of several dozen
\Msun. The best observed member of this class (and one of the very few known
altogether) is still SN\,2007bi \citep{galyam07bi}. This SN was hypothesised to
be the result of the PI explosion of a very massive star \citep{galyam07bi}.
This interpretation was based on finding a large ejected mass from both the
extended  light curve and the emitting mass inferred from the analysis of
emission lines in nebular spectra obtained at late time (more than one year
after maximum), but it has been challenged because the time of maximum may not
be as late as claimed by \citet{galyam07bi}, such that the light curve may not
be as broad as taht of PISN models, and because other SLSNe-I whose light curves
are not broad show similar spectra \citep{nicholl15}.

On the theoretical side, \citet{moriya10} suggested that SN\,2007bi may be
explained as the collapse of a very massive star. They were able to match the
light curve of SN\,2007bi with such a model, but did not address its
spectroscopic properties, so the suggestion remained unverified.  As various
groups have been developing physically based PISN explosion models that were not
available at the time of \citet{galyam07bi}, attempts have been made to compute
nebular spectra from these models. In particular, \citet{jerk16} showed that
PISN explosion models yield nebular spectra characterised by narrow emission
lines, which do not match those of SN\,2007bi, casting further doubt on the
nature of this and other SLSN-R. 

Determining the properties of PISN candidates is important in order to
understand the most massive stellar explosions and ultimately their impact on
the Universe. 

In this paper we compare PISN and massive core-collapse models in their ability
to match the observed late-time spectra of SLSN-R. Nebular-phase spectra are
ideally suited for the task of exploring the massive progenitors of SLSN-R, as
only at that epoch can we probe the entire ejecta. We extend the study to the
near-infrared (NIR), which turns out to be a critical spectral region. We use
models in their original version, but proceed to show how they need to be
modified in order to match the data. In a parallel paper \citep{moriya19} we
study the early phase of PISN candidates. 

In Section \ref{sec:data} we briefly discuss the data used for the comparison.
In Section \ref{sec:code} we describe our modelling code and strategy. In
Section \ref{sec:spectra} we present synthetic spectra computed using the
selected explosion models. In Section \ref{sec:pisn} we show how PISN models can
be modified to improve the fit to the data. In Section \ref{sec:mcc} we do the
same for massive core-collapse models, showing that they match the spectra of
SLSN-R very well, including the NIR, if some modification is introduced. In
Section \ref{sec:disc} we discuss our results. Section \ref{sec:concl} concludes
the paper.

\section{The Dataset}
\label{sec:data}

Very few nebular spectra of PISN candidates are available. Two spectra of
SN\,2007bi were published by \citet{galyam07bi}, and they remained the only
available nebular spectra of a SLSN-R for quite some time. Recently, as
observational efforts have increased, a few more PISN candidates have been
observed in the nebular phase. One of the best examples is SN\,2015bn 
\citep{nicholl15,jerk16}. The SN has good optical spectra but also (noisy) NIR
data, which are useful.  Here we focus on SN\,2007bi, the first PISN candidate.

A VLT-FORS spectrum of SN\,2007bi was obtained on 10 April 2008 (PI Mazzali) and
was published in \citet{galyam07bi}. The epoch of the spectrum is 367 rest-frame
days after a poorly defined light curve maximum. The spectrum has recently been
re-calibrated by \citet{jerk16}, who applied an improved subtraction of the host
galaxy. We use their calibration here. The newly established flux levels lead to
a new determination of the \Nifs\ mass. We adopt a distance modulus $\mu =
38.86$\,mags (corresponding to a distance of 592\,Mpc), and reddening $E(B-V) =
0.028$\,mags.

The spectrum displays a number of emission lines, mostly forbidden, of elements
ranging from oxygen to iron. The strongest optical lines can be identified as in
\citet{mazzali07gr} as, from blue to red: \CaII\ H\&K; \MgI] 4570; [\FeII] 5159,
5262, 5273, 5334, 5376 in a broad blend; \OI\ 5573; [\OI] 5577, 6300, 6363;
\NaI\,D; \CaII] 7291, 7324; \OI\ 7773.  \citet{mazzali07gr} provide also line
identifications in the near-infrared, which apply to our models below.

\section{Modelling approach}
\label{sec:code}

\begin{table*}
\caption{Model properties}
\label{table:models}
\centering
\begin{tabular}{lrccrrrrrrrr}
Model & \Mej & \KE & \KE/\Mej & Ni & C &  O & Ne & Mg & Si &  S & Ca \\
\hline
M100    &  94 & 4e52 & 0.45 & 3.1 & 5 & 46 &  4 &  4 & 18 &  9 &  1 \\
M110    & 100 & 5e52 & 0.52 & 9.1 & 5 & 38 &  4 &  4 & 21 & 12 &  2 \\
M30E30  &  33 & 3e52 & 0.91 & 4.5 & 1 & 14 &  1 &  1 &  5 &  3 &  1 \\
\hline
\end{tabular}
\end{table*}

We computed nebular-epoch spectra using our non-local thermodynamic equilibrium
(NLTE) code. The code was described in \citet{mazzali04eo}. It is based on the
method outlined by \citet{axelrod80}. The code computes the emission,
propagation and deposition of gamma rays and positrons from \Nifs\ and \Cofs\
decay using a Monte Carlo method, as outlined in \citet{cappellaro97}. The
energy deposited is used to excite and ionize the gas via collisions. The
heating is then balanced by cooling via emission in (mostly) forbidden lines.
The code assumes microscopic composition mixing within a radial density zone and
can implement ejecta clumping. This feature was not used in most of the ejecta,
but it was necessary to introduce mild clumping (filling factor $\sim 0.5$) in
the \Nifs-rich regions in order to suppress the presence of \FeIII\ via
recombination. This is common to all SN\,Ib/c nebular models
\citep{mazzali07gr}.  At the late epochs of the nebular spectra, \Nifs\ has
almost completely decayed, but cobalt does not show strong emission lines. Iron
is therefore the best tracer of the initial \Nifs\ content. In a PISN, which is
a thermonuclear explosion, as in SNe\,Ia, matching both Fe emission lines and
lines of elements that are not the product of radioactive decay makes it
possible to estimate both the heating (hence the \Nifs\ mass) and the cooling
(which is caused by all emitting species), and thus to evaluate simultaneously
both the \Nifs\ mass and the mass of the ejecta, at least in those regions that
are affected by collisional heating (\ie\ regions close to the location of
\Nifs). 

The profiles of the emission lines are indicative of the spacial distribution of
the emitting elements. Deviations from spherical symmetry can be appreciated
from the deviation of line profiles from the expected one-dimensional profile
\citep[\eg][]{mazzali05}. 

We used three different explosion models: two PISN models, M100 and M110, from
S.E. Woosley \citep[the names indicate the mass of the He core, see][and in this
case these models correspond to main-sequence stars of $\sim 200
\Msun$]{heger02}, as well as a massive ($30 \Msun$) core-collapse SN model,
originating from a $100 \Msun$ star model from \citet{umeda2008}. These
explosion models were selected because they synthesize $\sim 3 - 9$\,\Msun\ of
\Nifs, similar to the value estimated for SN\,2007bi if it is assumed that its
light curve is entirely powered by radioactivity \citep{moriya10}. All models
are one-dimensional, as is our spectrum synthesis code, and are characterised by
a ratio of kinetic energy of expansion and ejected mass of \KE/\Mej\,$\sim 0.5$
-- 1 [$10^{51}$ erg $\Msun^{-1}$]. A ratio of $\sim 1$ was shown to be adequate
for SLSN-I \citep{mazzali16}, which however show a great variety in \Mej\ and
M(\Nifs). PISN models uniquely predict kinetic energy and nucleosynthesis. As
for the massive collapse model, we explodesd it with an explosion energy
$3\times 10^{52}\,\mathrm{erg}$. This produces a \Nifs\ mass of $\approx
4.5\,\Msun$), similar to the value required to power SN\,2007bi
\citep{moriya10}. The ejecta structure and composition are based on the results
obtained by \citet{moriya10}. We call this massive core-collapse model M30E30.
Table \ref{table:models} lists the most important properties of the three
models.

\begin{figure*}
 \includegraphics[width=139mm]{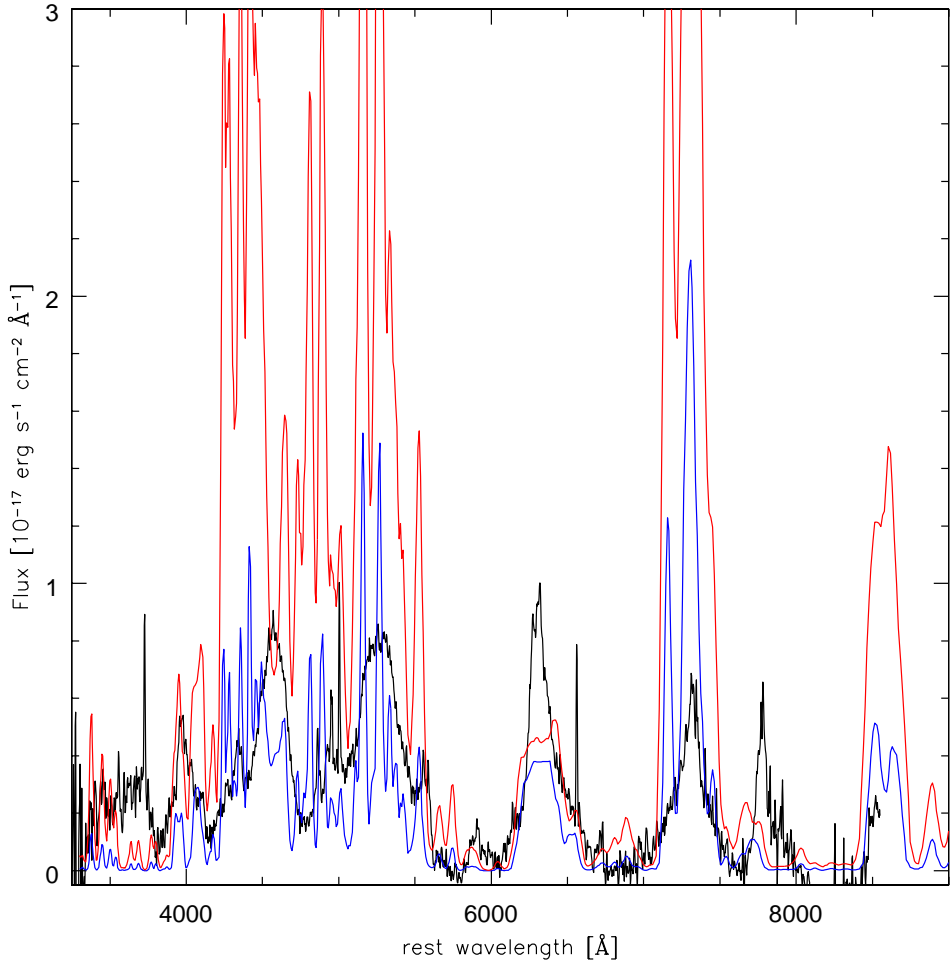}
\caption{Synthetic nebular spectrum of PISN models M100 (blue line) and M110 
(red line), compared to the nebular spectrum of SN\,2007bi (black line) obtained 
with VLT/FORS on 10 April 2008 (PI Mazzali) and published in \citet{galyam07bi}. 
The epoch of the spectrum is 367 rest-frame days after a poorly defined light 
curve maximum. As noted in the main text, PISN models yield spectra characterised 
by very narrow lines, which do not match the properties of the spectrum of 
SN\,2007bi. In particular, the models produce a flat-topped [\OI] 6300,6363 line 
rather than a sharply peaked emission, they fail to reproduce the \MgI] 4570 
line, and produce a forest of narrow [\FeII] lines instead of the broad emission 
seen in SN\,2007bi near 5250\,\AA.}
\label{fig:pisnneb}
\end{figure*}

First we computed synthetic spectra at the estimated epochs of the nebular
spectrum of SN\,2007bi using the density and abundance distribution of the
original models. The results are compared to the available spectra, and their
strengths and weaknesses are discussed. We then arbitrarily modified the models,
showing that simple modifications, which may be physically motivated, can lead
to much improved fits.

\section{Synthetic spectra}
\label{sec:spectra}

We computed synthetic nebular spectra for the three explosion models, two PISN
and one massive collapse. In order to make a comparison to observed spectra of
PISN candidates possible, we show the models at the distance and reddening of
SN\,2007bi. In our calculations we used a fiducial epoch of 435 days from
explosion following \citet{galyam07bi}, but see \citet{nicholl16} for a possibly
different estimate of the rise time of SN\,2007bi. Adopting a shorter rise time,
as is done, \eg in the companion paper presenting spectra for the early,
photospheric phase of a massive energetic collapse model and matching them to
SN\,2007bi \citep{moriya19} would lead to higher flux in the synthetic spectra
and therefore to a smaller estimate of the \Nifs\ mass required for SN\,2007bi.

\subsection{PISN models}

\begin{figure*}
 \includegraphics[width=139mm]{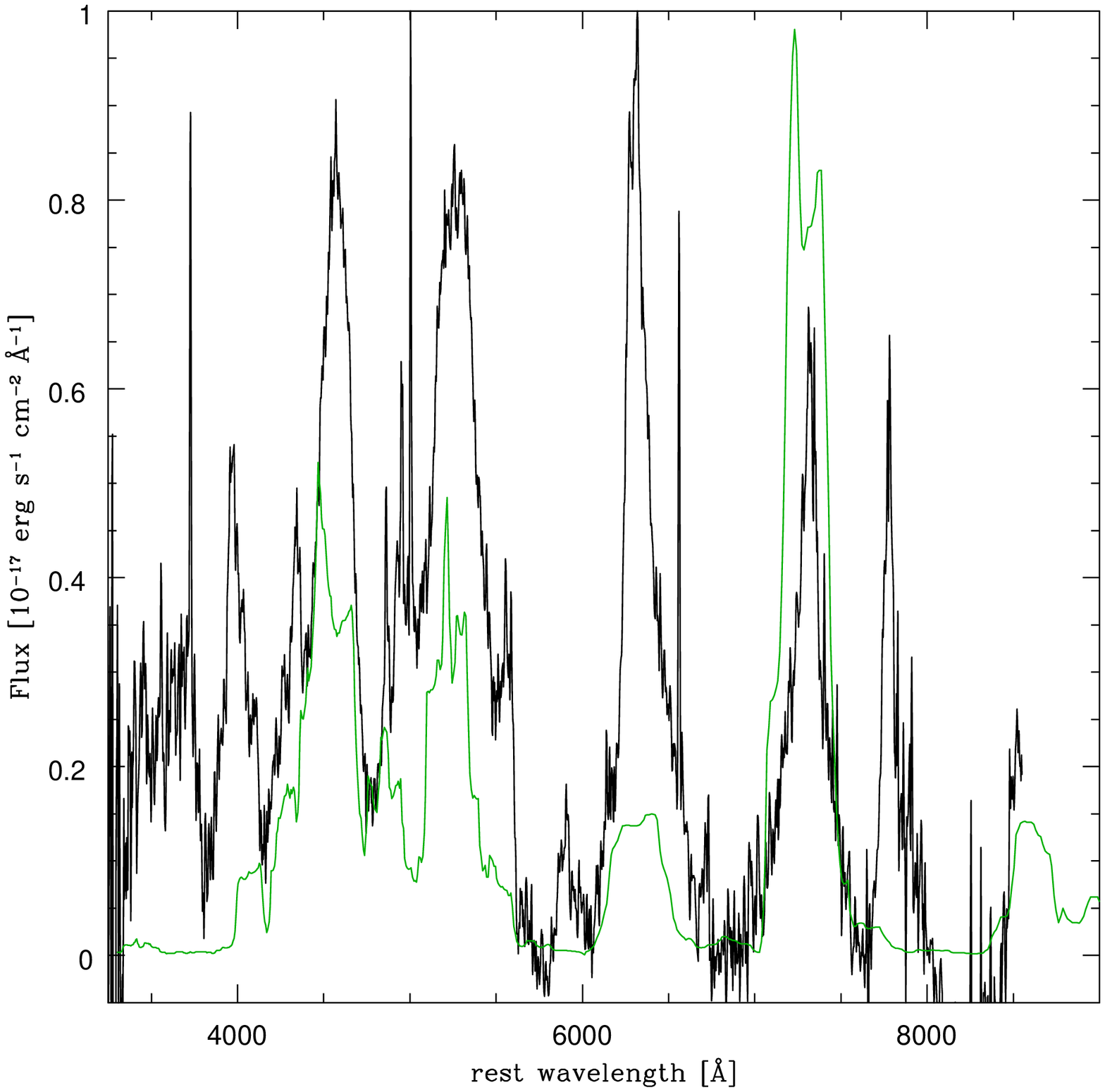}
\caption{Synthetic nebular spectrum of the MCC model M30E30 (green line), 
compared to the nebular spectrum of SN\,2007bi (black line).}
\label{fig:mcc}
\end{figure*}

The synthetic nebular spectra of the two PISN models, M100 and M110, are shown in
Fig. \ref{fig:pisnneb} and compared to the VLT nebular spectrum of SN\,2007bi. It
is immediately clear from Fig. \ref{fig:pisnneb} that the match between synthetic
spectra and observations is not satisfactory, primarily because the synthetic
spectra have narrow lines in most species, notably Fe. This is because in both
models almost all \Nifs\ is concentrated at the lowest velocities, below
3000\,\kms. This confirms the findings of \citet{jerk16}.  Additionally, model
M110 appears to be much too luminous for SN\,2007bi. The flux of model M100
appears to be closer to the observed spectrum of SN\,2007bi, but see caveat above
for the effect of a shorter risetime. 

On the positive side, many of the lines that are predicted to be strong in the
synthetic spectra of the two PISN models are also observed in SN\,2007bi. Most
notably, the [\OI] 6300, 6363 line is quite strong. The synthetic line is broad,
reflecting the distribution of oxygen, but - interestingly - it is truncated at
the lowest velocities and shows a flat top. This is the result of the fact that
no oxygen is present in the low-velocity, inner part of the PISN ejecta, where
\Nifs\ dominates, and so the low-velocity emission peak is suppressed.

\subsection{Massive core-collapse models}

The synthetic nebular spectrum obtained from model M30E30 is shown in Fig.
\ref{fig:mcc}. This spectrum shows much broader lines than the ones derived from
the PISN models. This is true in particular of the [\FeII] emission feature near
5200\,\AA, reflecting the broader distribution of \Nifs. The synthetic spectrum
shows strong lines of \MgI] 4570 and \CaII] 7291, 7324, as well as [\FeII]
lines.  The [\OI] 6300, 6363 line is present, but quite weak, and again it shows
a flat top, reflecting the lack of oxygen at low velocities in the model. Model
M30E30 has a similar \Nifs\ mass as M100 ($4.5$\,\Msun), but a much smaller
\Mej, 32.5\,\Msun, and consequently a much smaller oxygen mass, 14\,\Msun.
Overall, this spectrum seems better suited to match SN\,2007bi than the ones
based on PISN models,  although its flux is too low in most lines, suggesting
that the \Nifs\ mass in this particular model may be too small.  The flux is
somewhat weaker than that of the PISN model M100 in the optical, but PISN models
include significant amounts of IME (Si, S), which cool the gas and radiate
effectively in the NIR. This leads also to the weakness of the [\FeII] lines.
The NIR is therefore an essential part of the spectrum if we want to
differentiate among different models.

\subsection{The near-infrared}

\begin{figure*}
 \includegraphics[width=139mm]{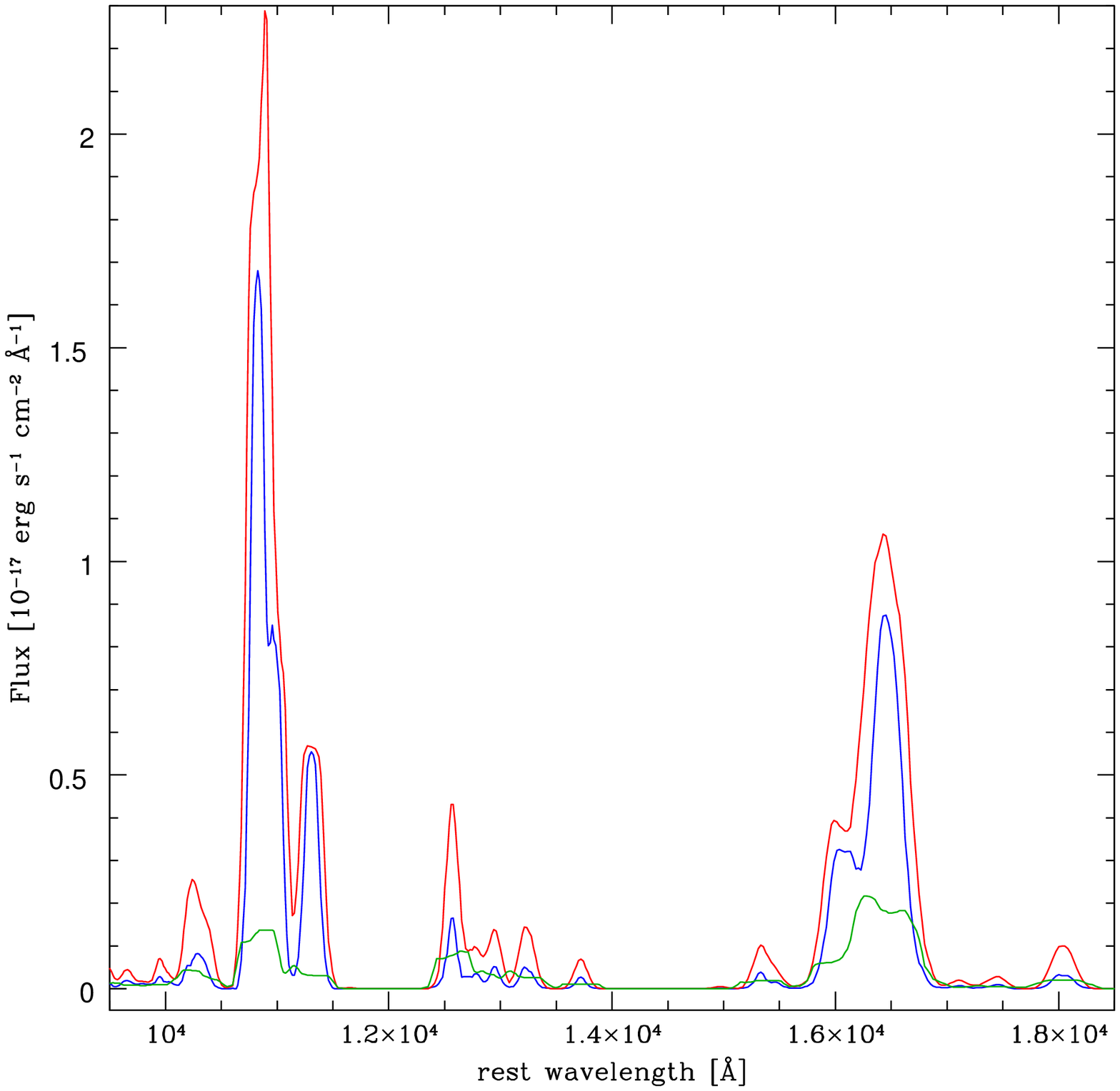}
\caption{Synthetic nebular spectra of the three models in the NIR: PISN models M100 (blue line) and M100 (red line); massive core-collapse model M30E30 (green line).}
\label{fig:nir_orig}
\end{figure*}

In Fig. \ref{fig:nir_orig} we show the predicted NIR behaviour of the three
models we have considered. The two PISN models emit strongly in a few IME lines:
[\SI] $1.08, 1.13, 1.15\mu$, [\SiI] $1.10, 1.60, 1.65\mu$, and show narrow
[\FeII] lines, \eg near $1.25\mu$.

The massive collapse model M30E30, on the other hand, has only weak flux in the
NIR. The IME content of this model is much smaller. Lacking IME cooling, which
takes place mostly via NIR emission lines, the MCC model cools much more
efficiently via optical emission lines such as [\OI] and [\FeII]. Therefore a
smaller oxygen mass, excited by a similar \Nifs\ mass, is sufficient to produce
an [\OI] 6300, 6363 line of strength comparable to that of the PISN models, even
though these have a much larger oxygen mass. 

Unfortunately, no NIR data are available for SN\,2007bi, making a direct
comparison impossible. Some data are available for SN\,2015bn, but the lines of
interest fall mostly in the atmospheric gaps because of the redshift of
SN\,2015bn \citep{jerk16}. We will investigate SN\,2015bn in separate work.

\section{Optimization of PISN models}
\label{sec:pisn}

PISN Model M100 has an adequate \Nifs\ mass to explain the light curve of
SN\,2007bi, but the flux in the synthetic nebular spectrum is not properly
distributed. The emission lines in SN\,2007bi are rather broad (FWHM $\sim
7000$\,\kms), while the synthetic spectra obtained from both PISN models are
characterised by narrow emission lines.

In order to obtain broader emission lines without changing the overall density
structure of the ejecta, one strategy is to mix the ejecta in abundance, without
modifying the density structure. This commonly used approach to mixing is rather
artificial, as a different density structure may result from abundance mixing,
and it is not supported by PISN calculations, but it is useful as a test to
highlight differences between the results of first principles, one-dimensional
PISN models and the spectrum of SN\,2007bi. 

\begin{figure*}
 \includegraphics[width=139mm]{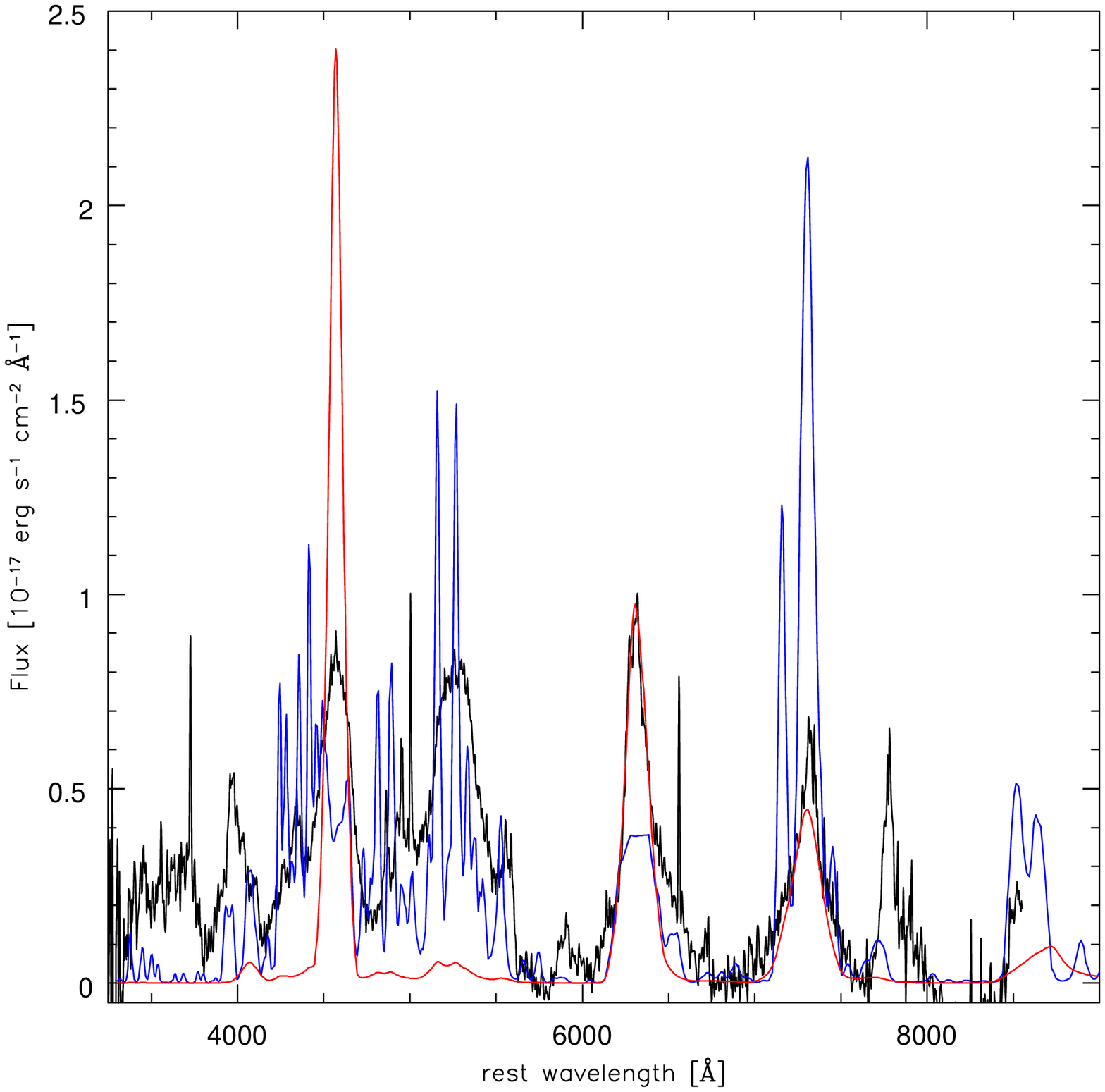}
\caption{Synthetic optical spectrum of the PISN model M100 with the original 
stratified abundances (blue line, same as in Fig. \ref{fig:pisnneb}) and with 
fully mixed abundances (red line), compared to the nebular spectrum of 
SN\,2007bi.}
\label{fig:mix}
\end{figure*}

The results of fully mixing model M100 in abundance are shown in Fig.
\ref{fig:mix}. It is interesting to note that when mixing is applied cooling
takes place in totally different lines than in the original model, which is very
sharply stratified in abundance. When microscopic abundance mixing is
introduced, at optical wavelengths light and intermediate mass (IME) elements
such as O, Mg, and Ca contribute most of the cooling, as they are now in close
contact with \Nifs\ heating.  Cooling happens mostly in three lines: \MgI]
4570\,\AA, [\OI] 6300, 6363\,\AA, and \CaII] 7291, 7324\,\AA, as shown by the
spectrum drawn in red in Fig. \ref{fig:mix}. The \MgI] line matches in
wavelength a strong emission in the spectrum of SN\,2007bi, suggesting that the
observed line is indeed \MgI] 4570\,\AA, and far exceeds the strength of the
observed feature. The [\OI] 6300, 6363\,\AA\ line matches the observed one
amazingly well, considering that no effort was made to fit the observations:
this suggests that the density structure of this model can describe SN\,2007bi
and that the ratio of radioactive heating and cooling via \OI\ is also adequate.
Finally, the \CaII] 7291, 7324\,\AA\ line also matches the observed line quite
well, the only shortcoming being that its flux is slightly smaller than the
observed one. 

Compared to the model with the original abundance stratification, the intensity
in the \MgI] line is stronger, because Mg is now present at low velocities,
close to the bulk of \Nifs; the \OI] line is stronger, in particular at the
smallest velocities, and it no longer has a flat-top profile, for the same
reason, while the \CaII] line is weaker, because calcium, an IME, was already
located next to \Nifs\ in the original, stratified model, but when full mixing
is introduced it competes with lighter elements, such as Mg and O, for
cooling.  

The species that is mostly affected by the increased cooling in light and
intermediate-mass elements is actually iron. The \FeII] emission lines are now
much weaker, as can be seen in Fig. \ref{fig:mix}.  On the other hand, abundance
mixing means that \Nifs\ is also present at all velocities, which leads to
broader, blended [\FeII] emission lines. \footnote{The same effect is
responsible for the broadening of Fe emission lines in SNe\,Ia
\citep{mazzali98}.}  In particular, a broad [\FeII] emission feature is now
present near 5200\,\AA.  This matches in wavelength an observed emission in
SN\,2007bi, but the synthetic feature is much too weak. This is because too much
of the cooling occurs via lighter elements than Fe, suggesting that the ratio of
\Nifs\ and ejected mass is too small in the PISN models. 

The match to the observed optical spectrum of SN\,2007bi is greatly improved by
introducing abundance mixing, but the synthetic spectrum is still far from
satisfactory in the optical, especially because of the weak iron emission.
Further improvements could be obtained by modifying the composition of the
ejecta. In particular, increasing the \Nifs\ mass would lead to more emission in
[\FeII] lines. Correspondingly, however, the O and Ca mass would have to be
reduced to avoid excessive cooling. This would reduce the overall mass to below
the nominal value of the model, and is therefore not an acceptable solution. The
abundance of Mg would also need to be drastically reduced. Another option would
be to introduce stable Fe to support the cooling. A model based on the density
of M100, with an increased \Nifs\ mass of $\sim 5 \Msun$ and containing some
$10\,\Msun$ of stable Fe would reproduce the observed spectrum reasonably well,
but the synthesis of large amounts of stable Fe is not foreseen in PISNe 
\citep{heger02}. 

\begin{figure*}
 \includegraphics[width=139mm]{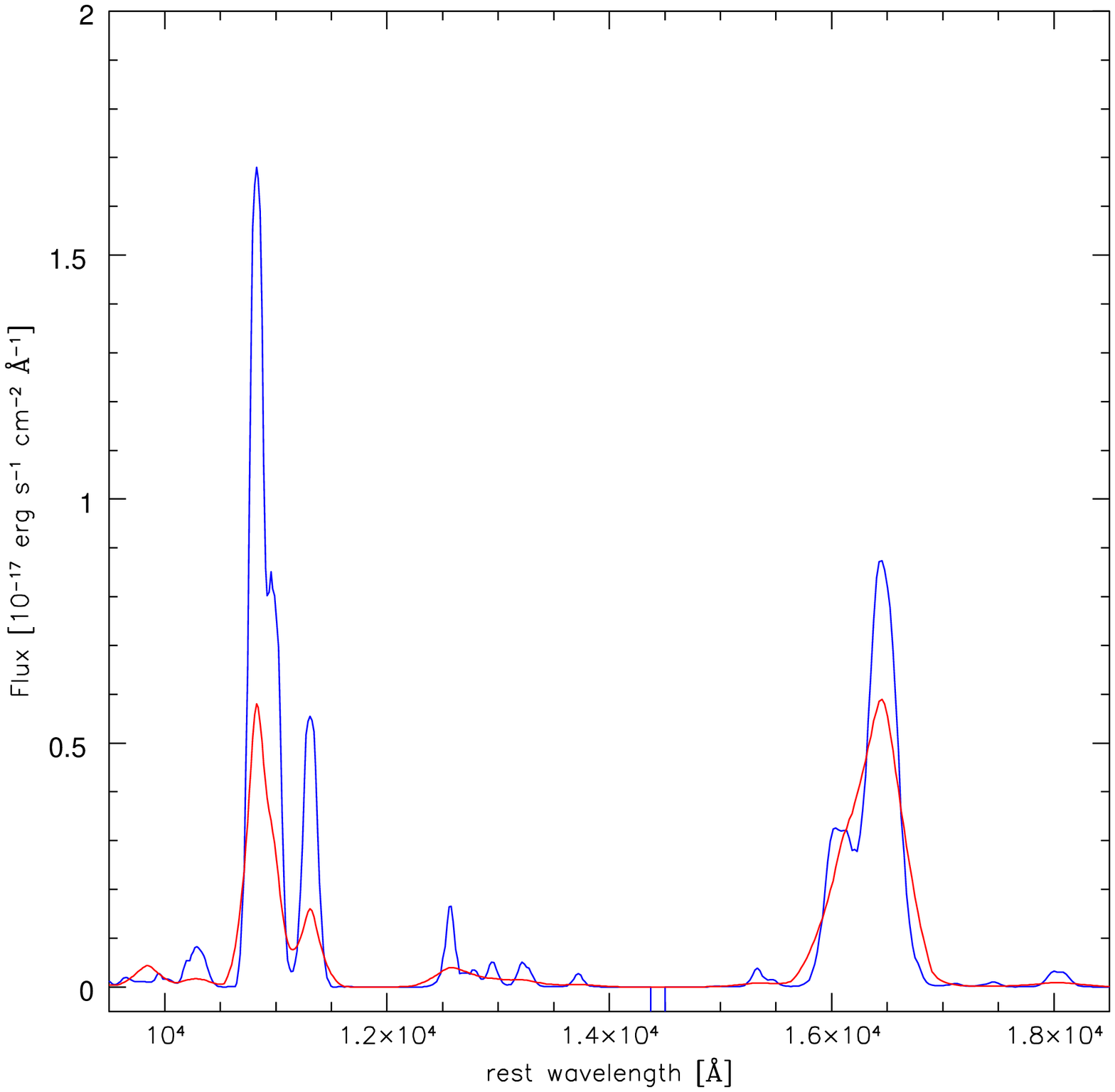}
\caption{Synthetic NIR spectrum of the PISN model M100 with the original 
stratified abundances (blue line, same as in Fig. \ref{fig:pisnneb}) and with 
fully mixed abundances (red line).}
\label{fig:mixnir}
\end{figure*}

The NIR spectrum of model M100 is also affected by the mixing, as shown in  Fig.
\ref{fig:mixnir}. As was the case for Ca in the optical, the emission in Si and
S, which are both IME, is reduced, as other, lighter species compete for cooling
the ejecta. Overall, more cooling occurs now in the optical and less in the NIR.
No NIR spectrum is available for SN\,2007bi.

The interesting results obtained from mixing the ejecta composition do not add
strong support to the PISN interpretation for SN\,2007bi. However, they suggest
that composition mixing, especially of \Nifs, and of lighter elements such as
oxygen and magnesium, is necessary to obtain emission lines similar to those
observed in SN\,2007bi. At the very least, some \Nifs\ must be present at
relatively high velocities, in order to broaden and blend the Fe lines, and some
oxygen must be present at low velocities in order to reproduce the observed
[\OI] 6300, 6363\,\AA\ emission profile.  Available one-dimensional PISN models
do not seem to possess the necessary characteristics. Multi-dimensional models
are unlikely to be drastically different, as the PISN event is essentially a
detonation, and large mixing is not expected in this case \citep{chen14}.

\section{Optimization of the massive core collapse model}
\label{sec:mcc}

As we have shown above, the massive core collapse (MCC) model produces a nebular
spectrum with features that resemble SN\,2007bi, including the broad [\FeII]
emission lines, a broad but weak [\OI] line and strong [\CaII] lines (see Fig.
\ref{fig:mcc}). Additionally, it has a weak NIR flux, which sets it apart from
PISN models (see Fig. \ref{fig:nir_orig}). Here we look at possible modifications
of this model that can improve its ability to reproduce the nebular spectrum of
SN\,2007bi. 

The \Nifs\ distribution in the MCC model M30E30 is not as narrowly confined in
velocity as it is in the PISN models, as shown by the width of the emission
lines. A clear shortcoming is however the truncation of the [\OI] line, which
reflects the lack of oxygen at the lowest velocities. This is not just a problem
of abundance mixing. Model M30E30 is one-dimensional, and it has a ``hole" in
density below $v_{\rm min} \approx 3700$\,\kms. This is by definition, as the
mass cut in the one-dimensional model is set at the corresponding mass to eject
an adequate \Nifs\ mass to account for the SN light curve,  while the mass below
this minimum velocity is assumed to be captured in the compact remnant.

However, this is not physically correct, as material with velocity just larger
than the escape velocity from the remnant ($v_{esc}$) should eventually be
ejected with a velocity close to zero. In addition, multi-dimensional motions
can move ejecta back to the inner layers. Therefore the ``hole" is likely to be
just the result of the one-dimensional explosion calculation. 

\begin{figure}
 \includegraphics[width=87mm]{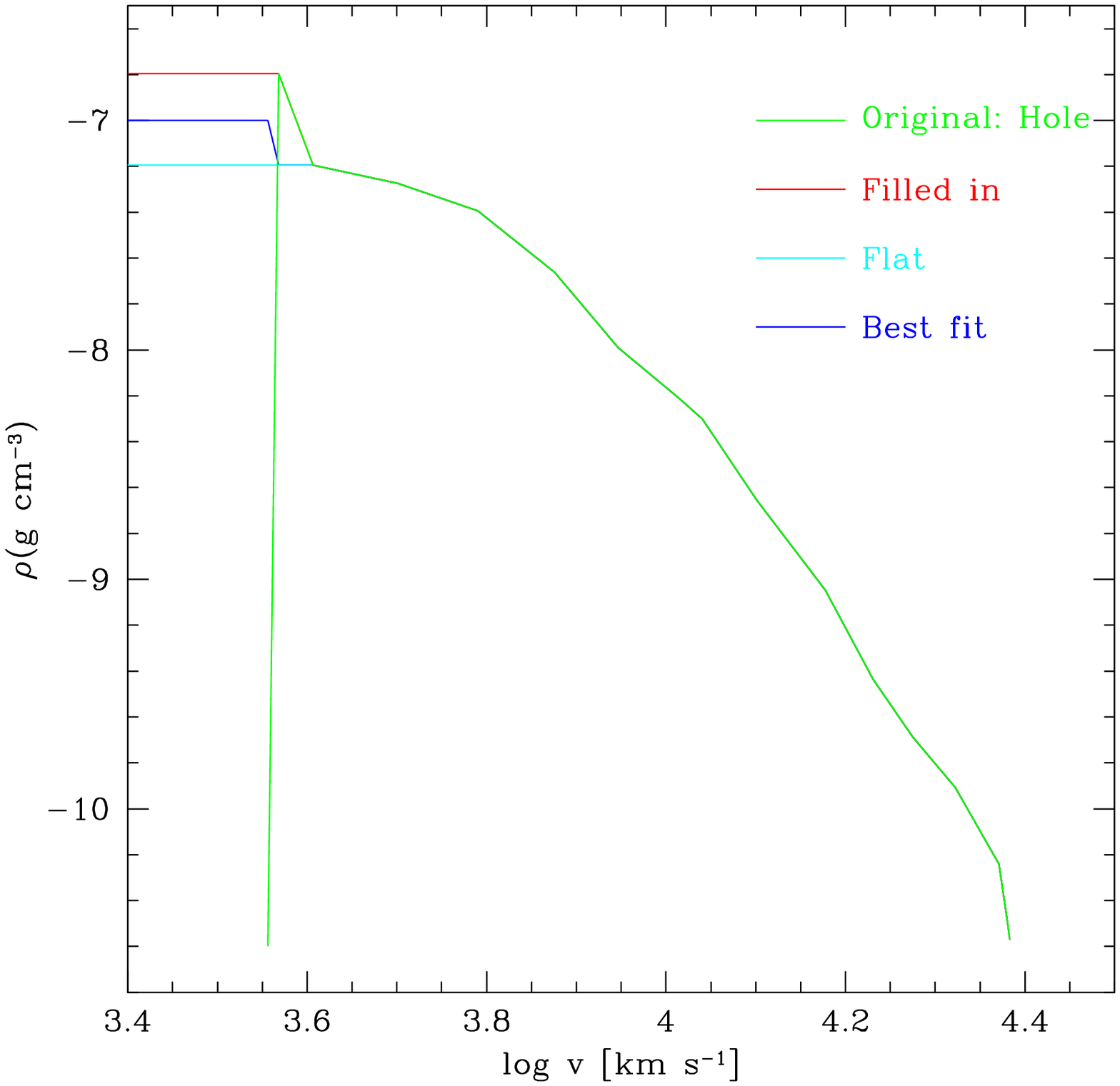}
\caption{Density profiles used in our calculations. The reference time for the 
density is one day after explosion.}
\label{fig:densities}
\end{figure}

The first modification we implemented was therefore to fill the inner hole with
additional ejecta. We looked at several possibilities, as regards both density
and abundances. The various density profiles we used are shown in Fig.
\ref{fig:densities}. It can be seen in that figure that the original density
profile rises smoothly to lower velocities but has a spike between 3700 and
3600\,\kms, followed by a very sharp drop. No mass is present below a velocity
of 3600\,\kms. This is the ``hole'' mentioned above.

\begin{figure*}
 \includegraphics[width=139mm]{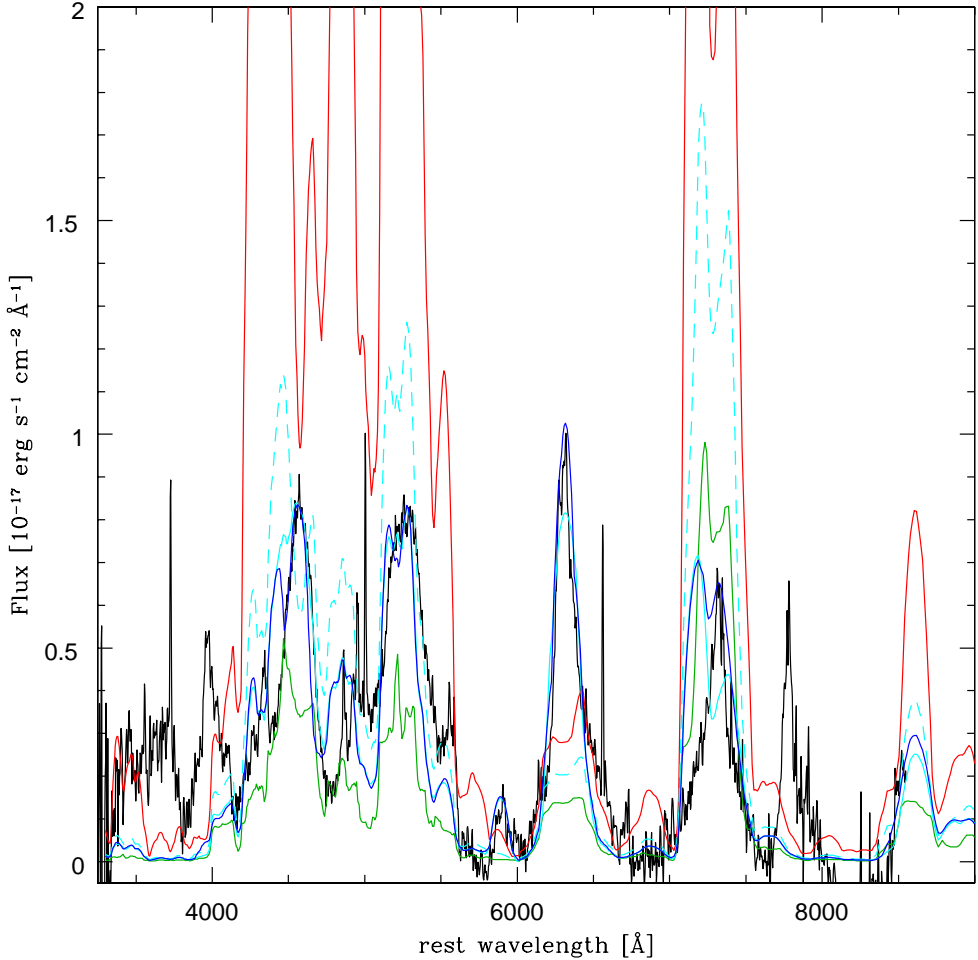}
\caption{Synthetic optical spectra of the massive core-collapse model M30E30 in 
the original version (green line, same as in Fig. \ref{fig:mcc}), with the inner 
hole filled-in (red), with constant (``Flat'') inner density and abundance 
either extended from the layers further out (dashed light blue line) or 
optimised (full light blue line), and a best-fit model optimised in both density 
and abundances (thick dark blue line), compared to the nebular spectrum of 
SN\,2007bi (thick black line).}
\label{fig:mccnohole}
\end{figure*}

In Fig. \ref{fig:mccnohole} the synthetic spectrum plotted in green is that
obtained from the original model. It was also shown in Fig. \ref{fig:mcc}. As we
discussed above, it shows broad lines, but is too weak in flux and has a
truncated [\OI] line. 

The first modification we implemented was therefore to fill in the hole with
constant density ejecta, starting from the onset of the hole in the original
model.  Initially, we applied to the newly filled innermost layers the same
abundances that characterise the original model in the layers just above the
hole. This implies that \Nifs\ is by far the dominant element in the newly
created innermost ejecta. Because of the high density in these innermost layers,
this model (shown in red in Figures \ref{fig:densities} and \ref{fig:mccnohole}
and termed ``filled in'') contains much too much \Nifs\ ($\approx 14.5\,\Msun$),
and the synthetic spectrum obtained from it is much too luminous. The ejected
mass also increases to \Mej$ = 42$\,\Msun. Increasing the \Nifs\ mass leads to a
larger luminosity, but does not solve the problem of the weakness of the [\OI]
line as oxygen is located too far from \Nifs\ to be heated by radioactive decay
products, which do not travel very far in the dense environment of the inner
core. Furthermore, because no oxygen has been added in the inner zones, the [\OI]
line still lacks the low velocity core. It is now double-peaked rather than flat.
This is not because the two individual components are separated, but mostly a
geometrical effect: the large amount of \Cofs\ decay energy available is carried
mostly by gamma rays. At late times gamma rays can partially penetrate into the
innermost oxygen layers and thermalise there. The thermalization process leads to
excitation of the levels from which the line is emitted. However, ejecta regions
lying further out are still shielded from gamma rays, and so collisional
excitation does not take place and neither does forbidden line emission. The
presence of an emitting oxygen shell results in a profile lacking central
(low-velocity) emission, such as what the models produce. Additionally, the fact
that decay from the upper level results in two emission lines separated by only
63\,AA\ also contributes to the observed red-to-blue peak ratio. We therefore
must abandon the idea that the inner layers contain \Nifs\ only. 

Next, we fill the inner hole but at the same time we eliminate the spike in
density between 3600 and 3700\,\kms. We call this density profile ``Flat''. We
still extend into the newly created innermost layers  the same abundances that
characterise the original model in the inner layers that have been preserved for
this particular density profile (now near 3700\,\kms). With this procedure,
\Nifs\ is still the dominant element in the innermost region. Its density
profile is plotted in light blue in Figure \ref{fig:densities}.  The model has
\Mej$ = 36$\,\Msun, and a \Nifs\ mass of 8.0\,\Msun.  The synthetic spectrum
computed using this density profile is shown as the dashed light blue line in
Figure \ref{fig:mccnohole}. The synthetic spectrum has too much flux in the
[\FeII], \MgI], and \CaII] lines, while the [\OI] line is too weak and still
shows a flat top. The oxygen mass is 15\,\Msun, but oxygen is mostly well
separated from \Nifs. 

As it is clear that the problem lies with the oxygen distribution, we then
tested mixing some oxygen down into the inner zones. To achieve this, we
modified the inner abundances of the ``Flat'' model from purely \Nifs\ to two
thirds \Nifs\ and one third oxygen, and also reduced the calcium abundance.  
This density profile still has an ejected mass \Mej$ = 36$\,\Msun, but the
\Nifs\ mass is now 5.6\,\Msun, less than in the ``Flat'' model but more than in
the original model M30E30, which could not reproduce the observed flux. The
total oxygen mass is now $\approx 21$\,\Msun. The resulting spectrum is plotted
as the light blue line in Figure \ref{fig:mccnohole}. This synthetic spectrum
has interesting features. First, the [\OI] line now has the correct shape, which
proves that there must be oxygen down to the lowest line-of-sight velocities.
The line is only slightly too weak in flux. Secondly, the [\FeII] lines produce
a smooth blend near 5200\,\AA, which reproduces quite closely that of
SN\,2007bi, confirming that a \Nifs\ mass of $\sim (5-6)$\,\Msun\ is required to
match the spectrum of SN\,2007bi. This \Nifs\ mass matches the mass estimated
from the light curve \citep{moriya10}. A slightly larger \Mej, leading to a
stronger [\OI] line, should reproduce the observations.

Finally, we produce an ad-hoc model to achieve a best fit. This is done by
fixing the density in the inner hole to a value intermediate among those used
above, while keeping the increased oxygen abundance. The density profile and the
corresponding synthetic spectrum of this ``best fit'' model are shown as dark
blue lines in Figures \ref{fig:densities} and \ref{fig:mccnohole}. Neither the
density profile nor the resulting synthetic spctrum are very different from the
``Flat'' model with low-velocity oxygen described just above.  The ``best fit''
model has \Mej$ = 38$\,\Msun, M(\Nifs)$= 5.5$\,\Msun, which guarantees a more
luminous spectrum, and an oxygen mass of $\approx 22$\,\Msun. The presence of
oxygen at low velocities forces significant cooling via [\OI] 6300, 6363\,\AA.
The synthetic spectrum matches most features quite nicely. In particular, the
[\FeII], \MgI], and [\OI] lines are quite well reproduced, although some \FeII]
lines are stronger than the observed ones and the \CaII\ H\&K emission is not
reproduced, assuming that is indeed the feature seen near 4000\,\AA\ in
SN\,2007bi. This may indicate that the density of Ca is higher than in our
model.

\begin{figure*}
 \includegraphics[width=139mm]{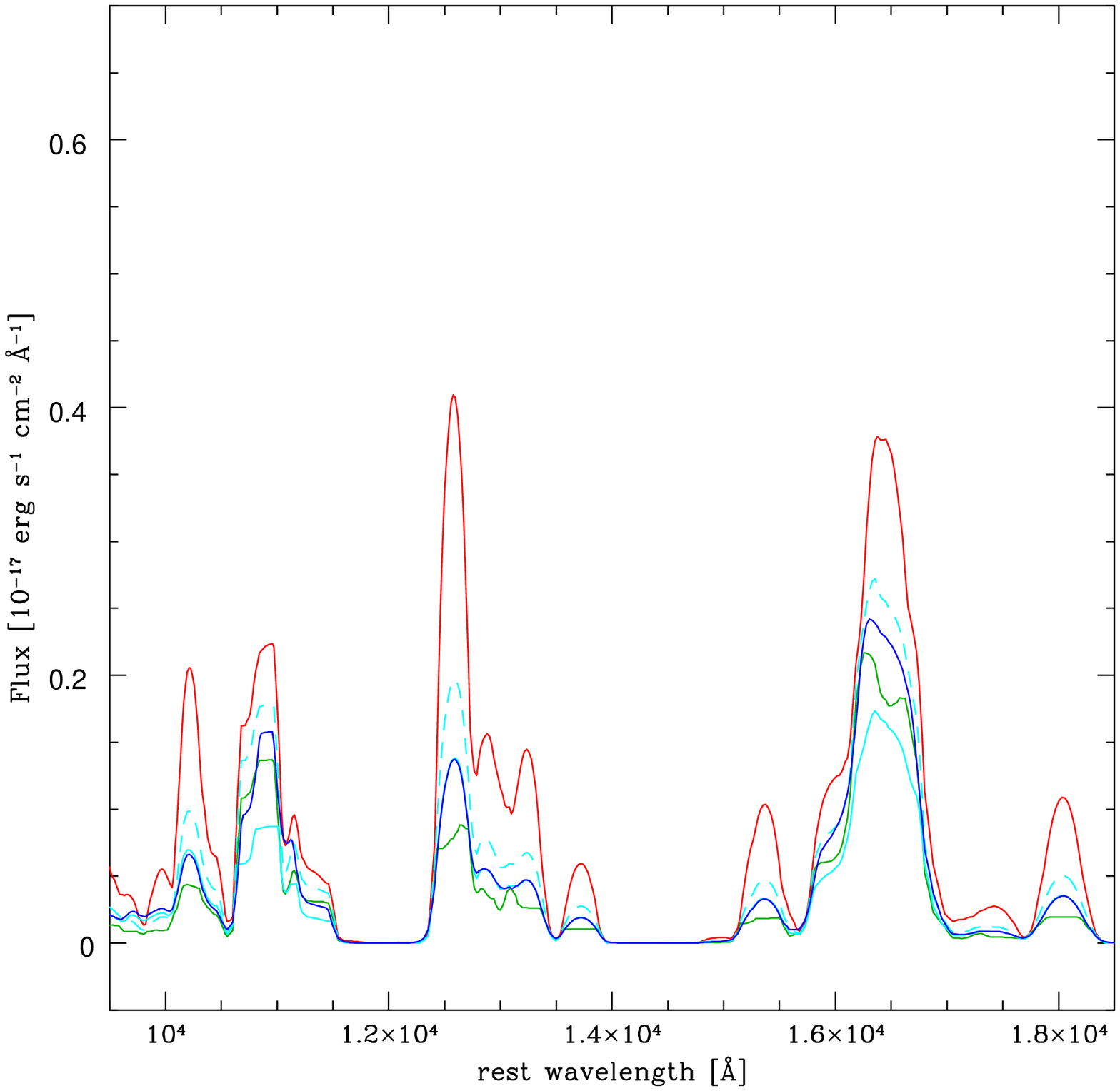}
\caption{Synthetic NIR spectra of the massive core-collapse model M30E30 in the 
original version (green line, same as in Fig. \ref{fig:mcc}), with the inner 
hole filled-in (red), with constant (``Flat'') inner density and abundance 
either extended from the layers further out (dashed light blue line) or 
optimised (full light blue line), and a best-fit model optimised in both density 
and abundances to match the optical spectrum of SN\,2007bi (thick dark blue 
line).}
\label{fig:mccnoholenir}
\end{figure*}

We turn now to the NIR. The four models we discussed are shown with the same
colour-coding in Fig. \ref{fig:mccnoholenir}. The synthetic spectrum with a
filled-in hole (red) has very strong NIR emission, a consequence of the very
large \Nifs\ mass, while the ``optimal" model, with only a slightly higher inner
density and enhanced oxygen content, is not very different from the synthetic
spectrum of the original model M30E30. This is because, as more heating goes
into oxygen at low velocity, the IME, which are located further out and dominate
emission at NIR wavelengths, are not much affected.

\section{Discussion}
\label{sec:disc}

Our nebular models address the question of the nature of SN\,2007bi. We showed
above that neither the PISN nor the MCC models produce good fits to SN\,2007bi
if they are used without modification. All the models we used are
one-dimensional calculations. PISN model He100 appears to be better suited to
match SN\,2007bi than the slightly more massive but much more luminous model
He110, mostly because of the amount of \Nifs\ synthesised. However, both models
produce very narrow emission lines, which do not resemble the nebular spectrum
of SN\,2007bi. The PISN model (we focussed on He100) can be improved by
thoroughly mixing the ejecta in abundance. This leads to a synthetic spectrum
with broader emission lines, in line with SN\,2007bi, but a theoretical
justification for this mixing is lacking. Furthermore, the very large content of
both oxygen and IME leads to strong emission lines of \OI, \MgI, \SiI, and \SI,
all at the expense of the \FeII\ lines, which are very weak, in sharp contrast
with the spectrum of SN\,2007bi. 
These problems might be alleviated if the PI explosion were grossly asymmetric
with e.g., the iron group coming out at high speed at high latitudes and the
oxygen ejected more slowly near the equator, as has been inferred for GRB/SNe
\citep{mazzali2001}. However, there is no compelling reason to think that PISN
are asymmetrical. Most of the angular momentum would be in the outer layers, and
only grossly differential rotation might produce an asymmetric bounce, but
magnetic torques are generally thought to suppress such strong differential
rotation. More work is needed to clarify thiese issues.

The MCC model has more intrinsic mixing, and so it naturally produces broader
lines, but in its original version it too has a number of problems. Its main
shortcoming is that the [\OI] 6300, 6363\,\AA, one of the strongest lines in the
spectrum of SN\,2007bi,  shows a flat top in the synthetic spectrum, while it is
quite sharp in SN\,2007bi. We showed that this is the direct consequence of the
presence of a central ``hole" in the density distribution. This ``hole" may be
the result of the one-dimensional calculation upon which the model is based.
Filling in the ``hole" leads to much improved fits ONLY if the innermost region
contains oxygen to a significant fraction (about one third by mass). This is a
challenge for one-dimensional models, and a strong indication that IF the MCC
model is adopted for SN\,2007bi the explosion must have been highly aspherical,
such that oxygen was synthesised at low velocities. This would not be unheard of
for massive core collapse explosions. There is ample evidence that the
explosions that give rise to luminous and energetic SNe\,Ic (sometimes
associated with Gamma-ray Bursts) are significantly aspherical. For the
prototypical GRB/SN, SN\,1998bw/GRB980425, a narrow [\OI] emission accompanied
by broad iron emissions also pointed to the presence of oxygen at low
velocities, possibly in a disc-like structure viewed close to face-on, such that
oxygen has a small line-of-sight velocity \citep{mazzali2001}. For other SNe,
this structure is indeed observed on its plane, leading to a strong [\OI] line
characterised by the presence of two widely separated peaks \citep{mazzali05}.
This morphology indicates quite an aspherical explosion. Whether this is
necessary also in SN\,2007bi is difficult to tell, as there are no other
indicators of asphericity. Yet, it has been pointed out that the spectra of this
SN resemble those of SN\,1998bw at late times \citep{jerk16}.

Another major discriminant between the two scenarios is the explosion \KE.  It
has been claimed that explosion \KE\ is a function of \Mej\ for SNe\,Ic at least
\citep[\eg][]{mazzali04aw}, and there is no doubt that the mass of the ejecta in
SN\,2007bi far exceeds that of any other well studied SN\,Ic, so it is quite
possible that the \KE\ is also high. Our calculations do suggest a high \KE, but
still retaining \KE/\Mej $\sim 1$. This is different from SNe\,Ic, where
\KE/\Mej\ grows more than linearly with mass. The origin of the \KE\ in these
explosions is unclear. In the PISN scenario thermonuclear burning of the massive
progenitor produces a large \KE, and does so roughly proportionally to the mass
\citep{heger02}, which is the appealing feature of this scenario. The
energization of core-collapse SNe remains an unsolved problem. For GRB/SNe the
collapsar model \citep{mcfw1999} could potentially lead to energetic explosions
via a neutrino-driven disc wind, but the link is not fully explored. In a normal
core-collapse event, neutrinos emitted in the neutronization of the nascent
neutron star may energise the ejecta, but it is unlikely that they can enforce
the very large energies observed in GRB/SNe, and possibly also the observed trend
of increasing \KE\ with increasing \Mej. Although SNe\,Ib/c are generally
consistent with being aspherical \citep[\eg][]{maeda08,tauben09}, it is unlikely
that this asphericity is caused by a jet, as GRB jets carry much less energy than
the SNe themselves \citep{mazzali14}, and they do not need to use much energy:
$\sim 10^{51}$\,erg is sufficient to penetrate a compact star
\citep{lazzati2013}, which is much less than the inferred \KE\ of GRB/SNe.  The
suggestion was made by \citet{mazzali14} that rapid energy injection from a
magnetar could energise the explosion and produce nucleosynthesis.\footnote{This
is not the scenario in which a magnetar powers the long-term SN light curve: the 
presence of strong [\FeII] emission lines in the nebular phase, constitutes good 
evidence that SN\,Ib/c light curves, including GRB/SNe, are driven by the 
radioactive decay of \Nifs\ \citep{mazzali2001}. These lines are a feature of
SN\,2007bi as well.} The energy of the massive collapse model we used is itself
compatible with the limiting magnetar energy ($\sim 3 \times 10^{52}$\,erg), so a
similar mechanism could be claimed. One difference between the GRB/SN situation
and that of the collapse of a very massive stellar core which could give rise to
events like SN\,2007bi may be that in the former case the proto-neutron star
(proto-magnetar) rapidly collapses to a black hole after depositing its
magneto-rotational energy in the surrounding stellar matter, powering the
energetic explosion and the synthesis of \Nifs. The black hole then powers jets
via accretion. These jets propagate in the already aspherically exploding stellar
material, and if they have enough energy to escape they eventually give rise to a
GRB. In the case of the massive collapse a black hole may not be formed, making
more core material available for the synthesis of large amounts of \Nifs\ which
then powers the SN light curve. Even if a black hole was formed, the exploding
stellar core in the MCC explosion that can be made to match SN\,2007bi is much
more massive than in typical GRB/SNe ($\sim 30$\,\Msun\ vs. $\sim 10$\,\Msun),
making it harder for a jet to escape.

Finally, the two scenarios, MCC and PISN, are likely to synthesize radically
different amounts of \Nifs, a feature that may be used to distinguish them. We
discuss the two scenarios in turn.

\subsection{\Nifs: Core Collapse Limits}

\begin{figure}
\includegraphics[width=0.58\textwidth]{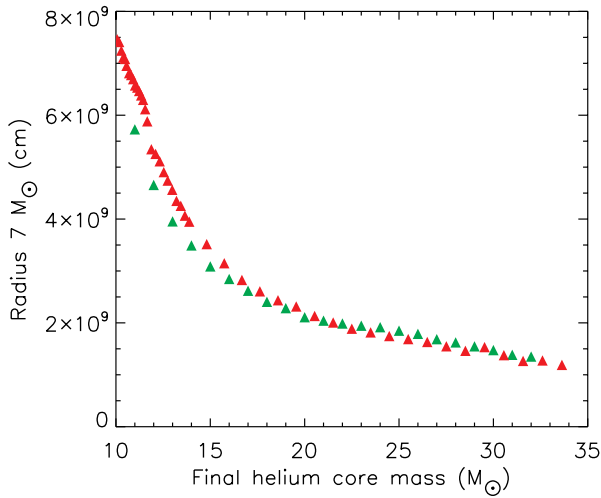}
\caption{Radius of the mass shell that encloses 7 \Msun\ as a
  function of the helium core mass of the presupernova star. Green
  points are for a series of helium cores evolved without mass loss. 
  Red points are for a series evolved including mass loss. Above
  35 \Msun, the cores are violently pulsationally unstable at oxygen
  ignition.  }
\label{fig:roft}
\end{figure}

According to \citet[][see their Fig. 26]{Woo02}, the shock temperature $T_s$ 
achieved at radius $R_s$ in a presupernova star is given, for a given point 
explosion of energy $E_{\rm SN}$, by the condition
\begin{equation}
E_{\rm SN} \ \approx \  \frac{4}{3} \pi R_s^3 a T_s^4.
\label{eq:esn}
\end{equation}
This equation basically states the dominance of radiation and pairs in
providing the pressure behind the shock and that most of the shocked volume 
is isobaric. This condition can be inverted to give the radius inside of 
which a given temperature, or greater, is achieved in the explosion,
\begin{equation}
R_{\rm s} \ \approx \ 1.0 \times 10^9 \ \left(\frac{5}{T_{\rm 9s}}\right)^{4/3} 
\, \left(\frac{E_{\rm SN}}{2 \times 10^{52} \ \rm erg}\right)^{1/3} \ {\rm cm},
\label{eq:roft}
\end{equation}
where $T_{\rm 9s}$ is the shock temperature in billions of degrees K. Outside
this radius only cooler temperatures are attained. A temperature of $5 \times
10^9$\,K is necessary to achieve nuclear statistical equilibrium on a
hydrodynamic time scale and produce iron-group elements \citep{Woo02}. The
explosion energy in any core collapse supernova is unlikely to exceed $2 \times
10^{52}$ erg, the maximum rotational energy for neutron stars without unusual
equations of state \citep[although see][for possibly a higher limit]{metzger05}.
Hence the production of iron in isotropic explosions is limited to regions with 
radius smaller than 10,000 km. 
This is actually a very generous limit on \Nifs\ since: 
a) it is unlikely that the full rotational energy of a cold neutron star can be
deposited in the time it takes the core to collapse and a shock wave to reach
10,000 km - less than a second; and b) not all the matter ejected from inside
this radius will be \Nifs. Roughly half will be helium from
photodisintegration. In calculations of $2 \times 10^{52}$ erg explosions in
massive helium cores, \citet{Woo17} found the synthesis of up to 2.7 \Msun\ of
\Nifs. In a 16 \Msun\ helium star, \citet{Nak01} found less than 1 \Msun\ of 
iron produced for explosion energies up to 10$^{53}$ erg.

Fig. \ref{fig:roft} shows the radius of the mass shell that encloses 7 \Msun\ in
a large number of helium stars evolved to the presupernova state, with and
without mass loss, using the \KEPLER\ code. Those evolved without mass loss
would be appropriate for single stars that do not lose their entire envelope as
supergiants. Those with mass loss are more representative of stars in binary
systems that lose their envelope shortly after red giant formation, which is
assumed here to happen at helium ignition. The models are extracted from a
survey in preparation by S.~E. Woosley. The value 7 \Msun\ is the sum of an
assumed 5 \Msun\ of ejecta plus 2 \Msun\ for the collapsed remnant. Only very
massive stars that leave such large remnants are capable of ejecting so much
iron.

The figure shows that only for the most massive cores is the critical radius of
10,000 km (Eq. \ref{eq:roft}) - almost - approached. Given that some of the mass
will not be \Nifs\ and that a supernova with \KE\,$\gtaprx 2 \times 10^{52}$ erg
s$^{-1}$ sustained for a second is unlikely, 5 \Msun\ of \Nifs\ ejected seems a
strong upper bound for core collapse SNe. Upper bounds of 2 - 3 \Msun\ may be
more realistic.

\subsection{\Nifs: PISN Limits}

The only supernova models that have been demonstrated to produce 5 \Msun\ of
\Nifs\ or more are PISNe \citep{heger02}. In fact, PISN models can produce up to
57\,\Msun\ of it. PISNe occur at zero-age main sequence (ZAMS) masses between
$\sim 140$ and $\sim 260$\,\Msun, but below $\sim 200$\,\Msun\ (\ie He core
masses between 65 and 100\,\Msun) they do not produce large amounts of \Nifs. 
The PISN mechanism requires helium cores from 100 to 133\,\Msun\ to produce more
than 5\,\Msun\ of \Nifs, and ZAMS masses roughly twice that. Such very massive
stars are quite rare. Producing them may not be so much of a problem as
maintaining that large mass until the star dies. One can dial the mass loss down
by reducing the metallicity in an uncertain scaling relation, but the existence
of such massive SN progenitors lacks observational support. Would $\eta$ Carinae
and similar massive stars maintain their original mass if their metallicity were
small? Metallicity only affects line and grain opacity, not global instabilities
caused by approaching the Eddington limit.  

If its light curve and spectra were powered exclusively by the radioactive decay
of \Nifs, as the strong Fe lines in the nebular spectra suggest, SN\,2007bi
synthesised $\sim 5$\,\Msun\ of it. A ZAMS mass of $\sim 200$\,\Msun,
corresponding to a He core mass of $\sim 100$\,\Msun, is required to produce so
much \Nifs\ \citep{heger02} as a PISN. IF SN\,2007bi and similar SLSNe are
PISNe, where are the other, fainter PISNe, produced at ZAMS masses between 140
and 200\,\Msun, and the more abundant pulsational PISNe produced by He cores
between $\sim 45$ and 65\,\Msun\ (\ie\ ZAMS mass between 100 and 140\,\Msun)
that the distribution of initial masses suggests should exist? Have they
possibly not been observed yet because of their intrinsic faintness combined
with their relative rarity? Because they neglect mass loss, the main sequence
masses assigned to a given final helium core mass by \citet{heger02} are lower bounds.

Perhaps the strongest evidence right now for a population of massive progenitors
are the black holes detected by LIGO \citep{Abbott16}, at least one of which
probably came from the collapse of a helium core over 35\,\Msun\ \citep{Woo16}.
It is a long way from 35 to 100\,\Msun\ He cores, above which PISNe are expected
to eject large amounts of \Nifs. PISNe leave no black hole signature, so they
cannot be traced in black hole mergers. The region of predicted low-luminosity
PISNe \citep{heger02} is yet unexplored. If SN\,2007bi was not a PISN but rather
a core-collapse SN, its progenitor (which we estimated here at a ZAMS mass of
100\,\Msun) would lie in this thus far unexplored mass range, implying that core
collapse SNe can happen at masses where direct black hole formation was thought
to be the only outcome.

\section{Conclusions}
\label{sec:concl}

There are pros and cons associated with both scenarios, MCC and PISN, when
trying to reproduce the observed properties of SN\,2007bi.  Can we distinguish
between them? 

Even if the two mechanisms could produce similar \Nifs\ mass (but see caveats
above), they will have different ejected mass and elemental abundances, in
particular as concerns IME. The two most interesting elements in this respect
are silicon and sulfur. \SI\ has an emission line in the optical
($\lambda\lambda 4507, 4589$), which is however blended with other lines, mostly
of Fe, and so it is not easy to use it as a proxy for abundance. The NIR is the
region where the two models are widely different. PISN models have a very large
IME mass in close proximity to \Nifs, which leads to very strong NIR emission
lines of [\SI] and [\SiI]. The MCC model, on the other hand, contains
fractionally much less IME, and its NIR flux is therefore much weaker. No NIR
data are available for SN\,2007bi. SN\,2015bn does have data \citep{jerk16} but
unfortunately, because of the redshift of the SN, the strongest IME emission
lines end up partially overlapping the atmospheric absorption bands. Yet,
SN\,2015bn does not appear to have very strong emission lines in the NIR, which
seems to favour the MCC option also for this SN. Whether SN\,2015bn may be
regarded as a PISN candidate based on the early-time data only is a different
question, which we will address in  separate work. Mixing or gross asphericity
may obviously affect both models. Given the evidence we obtained from nebular
spectroscopy, these issues require further study before conclusions can be
reached about the nature of this class of SLSNe.  

Three generic mechanisms for SN explosions of massive stars are known: neutrino
transport following iron core collapse; rotation, as in magnetar and collapsar
models; and pair-instability. If pair instability is ruled out because of lack of
mixing, and neutrinos are inadequate to produce an explosion with \KE$ \sim 2
\times 10^{52}$ erg, one is left with rotation, in some manifestation, as the
surviving explanation for SN\,2007bi.  This almost certainly implies some degree
of asymmetry in the explosion. The strong iron lines at late times indicate the
presence of at least some radioactivity in the explosion. The question is could
the light curve from \Nifs\ decay be supplemented to a large extent by magnetar
power, reducing the requirement for \Nifs? Could the approximate agreement of
iron abundance and the amount of \Nifs\ needed to explain the light curve be
fortuitous?  Given the assumption of our nebular analysis (\Nifs\ radioactive
powering, Fe coming from \Nifs\ decay), we feel that our results are solid.
Magnetar energy may be used to energise the explosion and aid the synthesis of
\Nifs, which then drive the light curve.

Another possibility, which is widely used in the SLSN literature, is to involve
circumstellar interaction. In the case of SN\,2007bi the circumstellar material
should be H- and He-free, as no characteristic emission lines are seen. The time
dependence of the interaction should also be tuned to mimic the effect of
radioactive decay of \Cofs, which also appears rather unlikely. 

No complete calculation of magnetar formation and \Nifs\ production exists, and
depositing a large fraction of the magnetar's energy in such a short time as to
raise the temperature of 5 \Msun\ of matter to $5 \times 10^9$ K seems - to us -
unlikely. However, if this was indeed possible many of the outstanding problems
with all energetic core-collapse SNe (energies in excess of $\sim 2 \times
10^{51}$ erg, large \Nifs\ production, aspherical distribution of ejecta) could
potentially be solved.

\section*{ACKNOWLEDGEMENTS}

We are grateful to Anders Jerkstrand for providing a newly subtracted spectrum of
SN\,2007bi.

This work was made possible by an NAOJ Visiting Joint Research grant, supported
by the Research Coordination Committee, National Astronomical Observatory of
Japan (NAOJ), National Institutes of Natural Sciences (NINS).

\end{document}